\documentclass[a4paper,11pt]{iopart}
\usepackage{cite}
\usepackage{iopams}
\usepackage{tikz}
\usepackage{subfigure}

\begin{document}

\title[Critical behavior of entropy production and learning rate]{Critical behavior of entropy production and learning rate: Ising model with an oscillating field}
\author{Yirui Zhang$^{1,2}$ and Andre C Barato$^3$}
\address{
$^1$II. Institut f\"ur Theoretische Physik, Universit\"at Stuttgart\\ 
Stuttgart 70550, Germany\\
$^2$ Institute of Physical Chemistry, Polish Academy of Sciences, Kasprzaka
44/52, 01224 Warsaw, Poland\\
$^3$ Max Planck Institute for the Physics of Complex Systems,\\
N\"othnitzer Str. 38, 01187 Dresden, Germany\\
}

\def\ex#1{\langle #1 \rangle}
\begin{abstract}
We study the critical behavior of the entropy production of the Ising model subject to a magnetic field that
oscillates in time. The mean-field model displays a phase transition that can be either first or second-order, depending 
on the amplitude of the field and on the frequency of oscillation. Within this approximation the 
entropy production rate is shown to have a discontinuity when the transition is first-order and to be continuous, 
with a jump in its first derivative, if the transition is second-order. In two dimensions, we find with numerical simulations 
that the critical behavior of the entropy production rate is the same, independent of the frequency and amplitude of the field. 
Its first derivative has a logarithmic divergence at the critical point. This result is in agreement with the lack of a first-order 
phase transition in two dimensions. We analyze a model with a field that changes at stochastic time-intervals between two values. 
This model allows for an informational theoretic interpretation, with the system as a sensor that follows the external field. 
We calculate numerically a lower bound on the learning rate, which quantifies how much information the system obtains about 
the field. Its first derivative with respect to temperature is found to have a jump at the critical point. 
\end{abstract}

\section{Introduction}

Nonequilibrium systems can display phase transitions, with a number of well understood universality classes \cite{hinr00,henk08,odor08}.
Some features not observed in equilibrium systems can occur if detailed balance is not fulfilled, e.g., correlations with a power law decay far 
from criticality and phase transitions with short range interactions in one dimension. Examples of nonequilibrium phase transitions include boundary induced phase transitions \cite{krug91,blyt07}, 
phase transitions into absorbing states \cite{hinr00}, real space condensation \cite{evan05}, and a transition to collective motion in active systems \cite{tama95,solo15}.   

The production of entropy is a signature of systems out of equilibrium. This entropy production can be defined 
for many nonequilibrium models, which can be related to quite different phenomena. Such a feature makes the 
investigation of the critical behavior of entropy production appealing. It is intriguing to wonder 
whether nonequilibrium phase transitions can be classified with respect to the critical behavior of the entropy production.

The critical behavior of the entropy production rate has been analyzed in the majority vote model \cite{croc05}, in a two-dimensional Ising model in contact 
with two heat baths \cite{deol11,tome12}, and in a model for nonequilibrium wetting \cite{bara12}. For the first two models, the first derivative of the 
entropy production rate with respect to the control parameter was found to diverge at the critical point. For the third model the first derivative of the entropy production rate
was found to be discontinuous at criticality. This discontinuity in the first derivative was also found within a mean-field approximation of the first model \cite{croc05}.
Furthermore, the entropy production rate of a model for population dynamics with a non-equilibrium phase transition has also been analyzed in \cite{andr10}.

In this paper we investigate the critical behavior of the entropy production rate  in an Ising model driven by 
a magnetic field that oscillates deterministically in time. This model displays a phase transition characterized by an order 
parameter given by the magnetization integrated over a period \cite{tome90,chak99}. For the mean-field model, the phase 
transition can be either first or second-order, depending on the field amplitude and frequency. For the two dimensional
model, the same kind of phase diagram with first and second-order phase transitions has been observed \cite{chak99}. However, a more careful 
numerical analysis indicates that for the two dimensional model the transition is always second-order \cite{korn02}, with critical 
exponents compatible with the Ising universality class \cite{korn01,buen08,fuji01}.
In contrast to the models listed in the previous paragraph, this model is driven by an external protocol 
and, therefore, reaches a periodic steady state \cite{sini07a,raha08,astu11,raz16}.

We analyze both a mean-field Ising model where all spins interact with each other and a two-dimensional Ising model with nearest neighbors interactions. 
Within the mean-field approximation, the entropy production rate is found to have a kink at the critical point for a second-order phase transition 
and is found to be discontinuous at criticality for a first-order phase transition. For the two-dimensional model 
the first derivative of the entropy production rate is found to diverge at the critical point, independent of the the frequency and amplitude of the field, which is
in agreement with the absence of a first-order phase transition in two dimensions.  

An Ising model with a field that changes at stochastic time-intervals  between two values is also considered.
The critical behavior of the entropy production rate does not change in relation to the one observed 
in the model with a deterministic field. However, this model allows for a further 
perspective related to the relation between information and thermodynamics \cite{saga12,horo13,mand12,bara14,bara14b,hart14,horo14,parr15,bara14a,hart16}.
The field can be seen as a signal and the system as a sensor that follows this signal. An information theoretic observable that quantifies the rate at which  
the system obtains information about the field is the learning rate \cite{bara14a,hart16}. This learning rate appears in a second law inequality for bipartite systems \cite{hart14,horo14}, 
being bounded by the thermodynamic entropy production rate that quantifies heat dissipation \cite{bara14a}. 

We study the critical behavior of the learning rate. Specifically, we introduce a lower bound on the learning rate that can be calculated in 
numerical simulations. Its first derivative with respect to the temperature is found to be discontinuous at the critical point.   

The paper is organized in the following way. In Sec. \ref{sec2} we calculate the entropy production rate for the mean-field model. The two-dimensional model
with a deterministic field is analyzed in Sec. \ref{sec3}. In Sec. \ref{sec4} we introduce the model with a field that changes at stochastic time-intervals 
and investigate the critical behavior of the learning rate. We conclude in Sec. \ref{sec5}.

\section{Mean-field approximation}
\label{sec2}

\subsection{Model definition and phase diagram}
We consider a Curie-Weiss mean-field Ising model of $L^2$ spins $s_i=\pm 1$ that is subjected to a time-dependent external field of strength $h(t)$. The time-dependent Hamiltonian
is given by
\begin{equation}
H(t)\equiv -\frac{J}{L^2}\sum_{ ij} s_is_j-h(t)\sum_i{s_i},
\end{equation}
where the first term on the right hand side involves a sum over all spins. The size-dependent pre-factor in this first term makes the Hamiltonian extensive. The external field 
varies periodically with time as
\begin{equation}
h(t)\equiv h_0 \cos(\omega t),
\label{eqfield}
\end{equation}
i.e., the field oscillates with an amplitude $h_0$ and a frequency $\omega$.

We consider a model with Markovian dynamics. The transition rate from configuration $(s_1,\ldots,s_i,\ldots,s_{L^2})$ 
to configuration $(s_1,\ldots,-s_i,\ldots,s_{L^2})$ is denoted by $w_i(s_i)$. The probability of an state $\mathbf{s}=(s_1,\ldots,s_i,\ldots,s_{L^2})$
at time $t$ is  denoted $P_\mathbf{s}(t)$. The average magnetization at time $t$ is given by 
\begin{equation}
m(t)\equiv \sum_\mathbf{s} s_iP_\mathbf{s}(t),
\end{equation}
where the above definition is independent of $i$ due to homogeneity.

Even though the model does not reach an equilibrium state due to the periodic variation of the external protocol, the transition rates at a fixed 
time $t$ fulfill the detailed balance condition 
\begin{equation}
\frac{w_i(s_i)}{w_i(-s_i)}= \textrm{e}^{-2 h_i(t)s_i/T},
\end{equation}
where $T$ is the temperature, Boltzmann constant is $k_B=1$ throughout, and 
\begin{equation}
h_i(t)\equiv \frac{J}{L^2}\sum_js_j+h(t).
\end{equation}
Assuming Glauber transition rates, i.e.,
\begin{equation}
w_i(s_i)= \frac{1}{2\tau}[1-s_i \tanh( h_i/T)],
\end{equation}
where $\tau$ sets the time-scale for a spin flip, it is possible to show that the magnetization follows the equation \cite{tome90} 
\begin{equation}
\tau \frac{dm}{dt}= -m(t)+\langle\tanh[h_i(t)/T]\rangle.
\label{eqmag1}
\end{equation}
In the thermodynamic limit $L\to \infty$ this equation is simplified by the relation $\sum_js_j/L^2\to m(t)$. 
Hence, in this limit, Eq. (\ref{eqmag1}) becomes 
\begin{equation}
\tau \frac{dm}{dt}= -m(t)+\tanh[Jm(t)/T+h(t)/T].
\label{eqmag2}
\end{equation}

\begin{figure}
\centering\includegraphics[width=90mm]{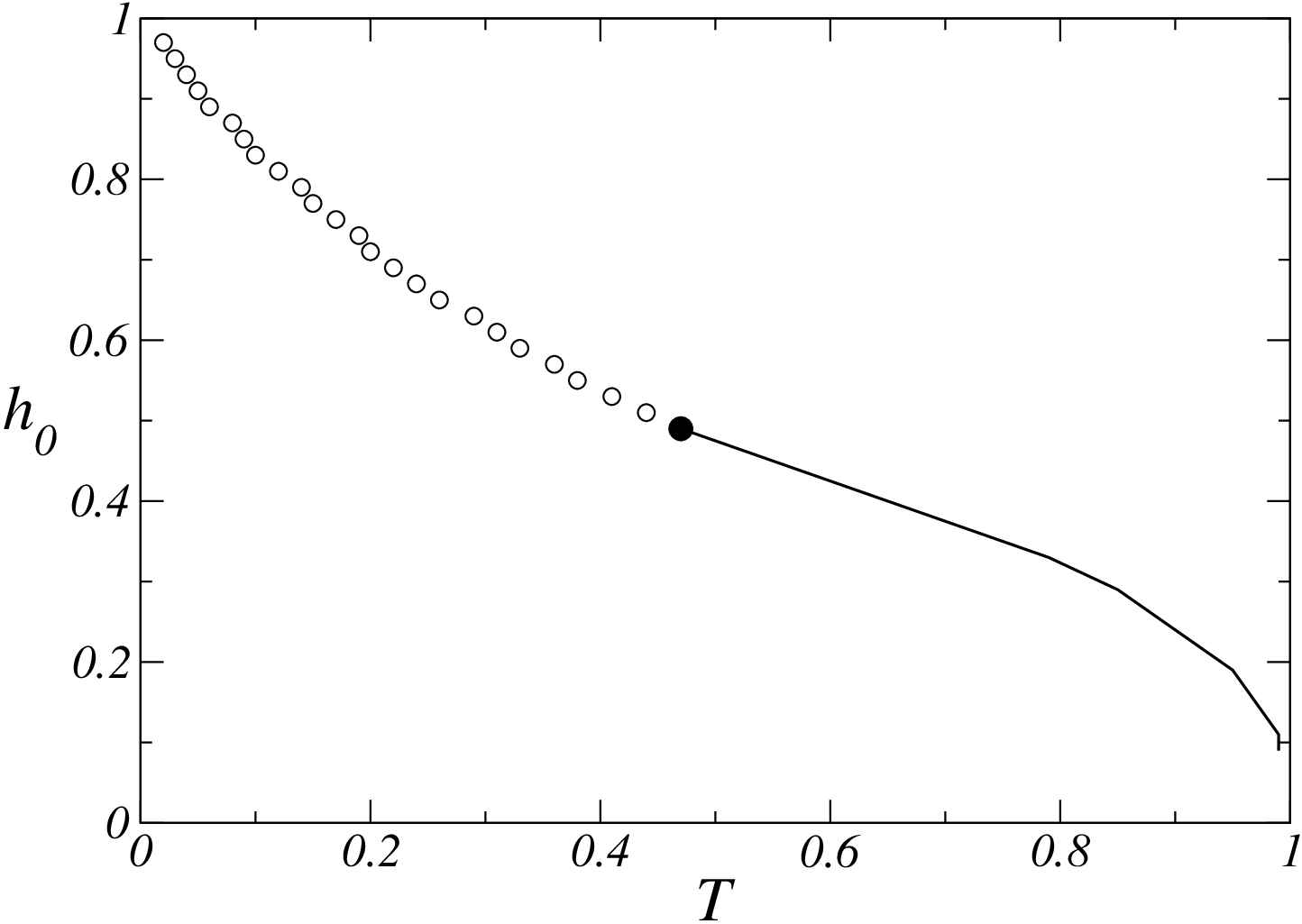}
\vspace{-3mm}
\caption{Phase diagram of the mean-field model. The parameters 
in Eq. (\ref{eqmag2}) are set to $J=\tau=1$ and $\omega= 0.2\pi$. In the region marked by the dots the transition is first-order and
in the solid line the transition is second-order. Above (below) the critical line the order parameter is $M=0$ ($M\neq0$).}
\label{fig1}
\end{figure}
The solution of Eq. (\ref{eqmag2}) reaches a periodic steady state, i.e., $m(t)=m(t+2\pi/\omega)$, that is independent of the initial 
condition. 

This model has a phase transition at a critical temperature $T_c$ that depends on the amplitude of the field $h_0$ and the 
frequency $\omega$. The order parameter of this transition is the magnetization integrated over a period in the periodic steady state  
\begin{equation}
M\equiv \frac{\omega}{2\pi}\int_{0}^{2\pi/\omega}m(t)dt.
\label{magdef}
\end{equation}
Below (above) the critical temperature the magnetization is $M\neq0$ ($M=0$). A phase diagram obtained with numerical integration of Eq. (\ref{eqmag2}) is shown in 
Fig. \ref{fig1}. This phase diagram has been obtained in \cite{tome90} and is shown here for illustrative purposes. 
Depending on $h_0$ and $\omega$ the phase transition can be first-order with a discontinuity in $M$, or second-order.

\subsection{Heat and Work}

Taking the time derivative of the internal energy per spin $u(t)\equiv H/L^2$, we obtain
\begin{equation}
\frac{du}{dt}= \frac{du}{dm}\frac{dm}{dt}+\frac{du}{dh}\frac{dh}{dt}. 
\end{equation}
Following the standard definition of work in stochastic thermodynamics \cite{seif12}, we identify 
the rate of work done on the system as
\begin{equation}
\dot{w}(t)\equiv \frac{du}{dh}\frac{dh}{dt}.
\label{eqmfw}
\end{equation}
The expression for the dissipated heat follows from the first law 
\begin{equation}
\dot{q}(t)\equiv \dot{w}(t)-\frac{du}{dt}= -\frac{du}{dm}\frac{dm}{dt}. 
\end{equation}
Since $u(t)$ is periodic, we obtain 
\begin{equation}
\int_0^{2\pi/\omega}\dot{w}(t)dt= \int_0^{2\pi/\omega}\dot{q}(t)dt, 
\label{eqmfwq}
\end{equation}
i.e., the average work done on the system in one period equals the average dissipated heat in one period.
The entropy production rate in the periodic steady state is defined as
\begin{equation}
\sigma\equiv \frac{1}{T}\frac{\omega}{2\pi}\int_{0}^{2\pi/\omega}\dot{q}(t)dt=\frac{1}{T}\frac{\omega}{2\pi}\int_{0}^{2\pi/\omega}\dot{w}(t)dt,
\label{eqmfsig}
\end{equation}
where we used the first law (\ref{eqmfwq}) in the second equality.

\begin{figure}
\subfigure[]{\includegraphics[width=75mm]{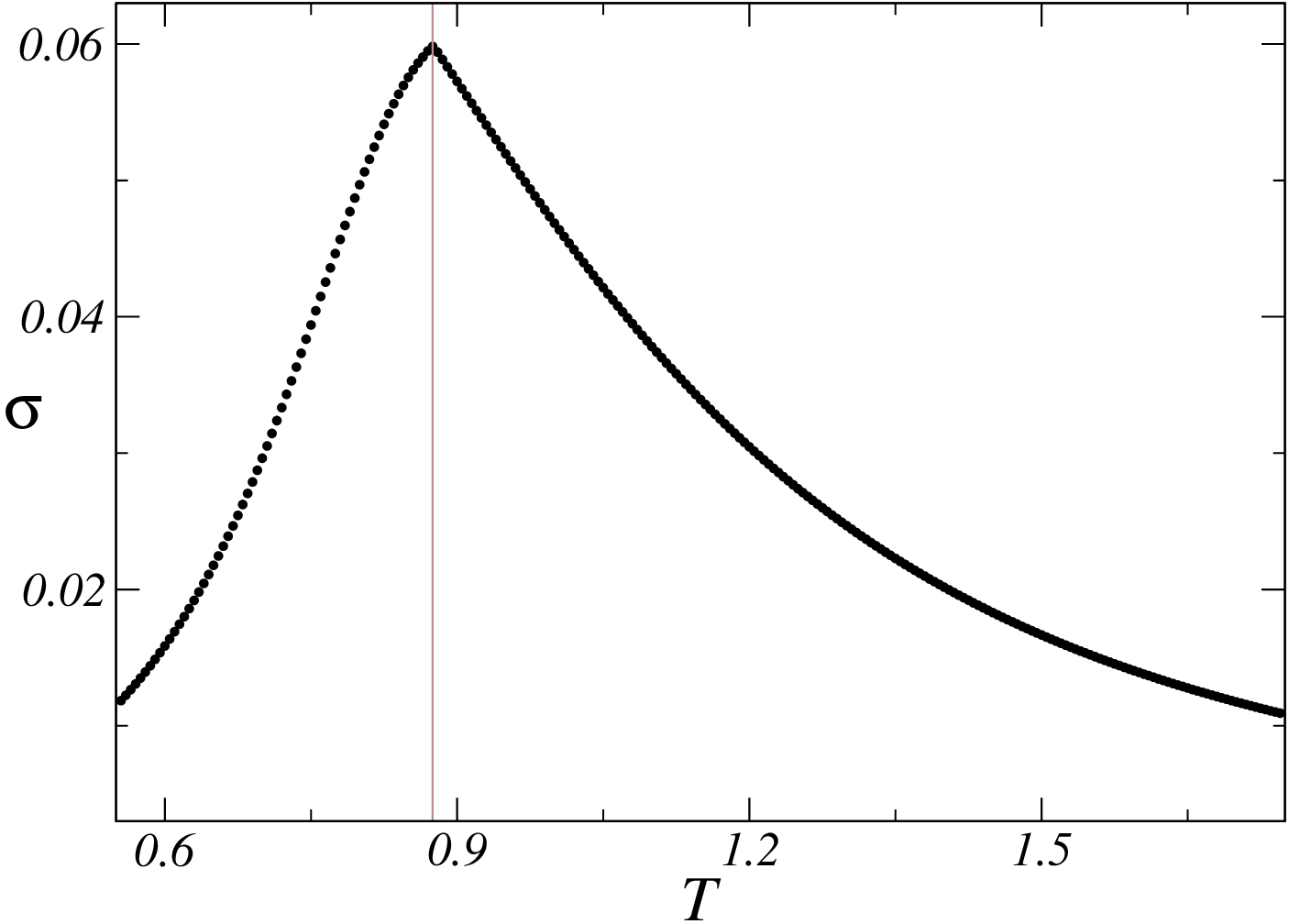}\label{fig2a}}
\subfigure[]{\includegraphics[width=75mm]{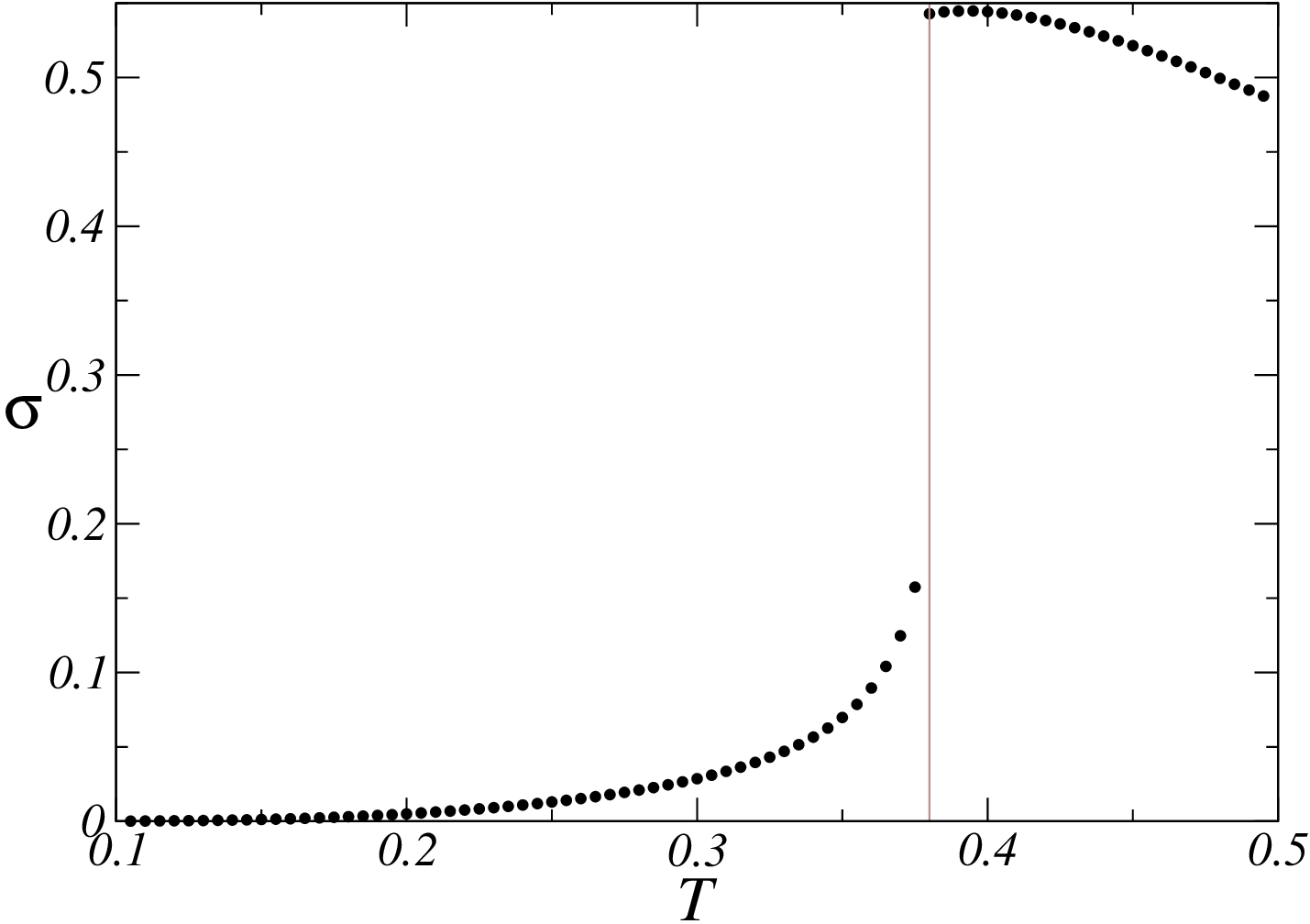}\label{fig2b}}
\vspace{-3mm}
\caption{Critical behavior of the entropy production rate for the mean-field model. The vertical lines mark the critical points. The parameters 
in Eq. (\ref{eqmag2}) are set to $J=\tau=1$ and $\omega= 0.2\pi$. (a) $h_0=0.25$, corresponding to a second-order phase transition. (b) 
$h_0=0.55$, corresponding to a first-order phase transition. 
}
\label{fig2}
\end{figure}

We solved equation (\ref{eqmag2}) numerically and calculated the entropy production rate $\sigma$ with Eqs. (\ref{eqmfw}) and (\ref{eqmfsig}). The results 
are shown in Fig. \ref{fig2}, where we plot $\sigma$ as a function of the temperature $T$ for two different values of the amplitude $h_0$ and fixed frequency $\omega$. 
If $h_0$ is such that the phase transition is second-order, the entropy production rate has a kink at the critical point, indicating that the first derivative of $\sigma$
with respect to $T$ has a discontinuity at criticality. If the phase transition is first-order, the entropy production rate itself is discontinuous  at the critical point. 
Hence, within the mean-field model the critical behavior of the entropy production rate is different in the two different regions of the phase diagram. 
We note that a discontinuity in the first derivative of $\sigma$ with respect to the control parameter has also been
observed in a mean-field approximation of the majority vote model \cite{croc05} and in a model for nonequilibrium wetting \cite{bara12}.

\section{Two-dimensional Ising model}
\label{sec3}

\subsection{Model definition}

\begin{figure}
\subfigure[]{\includegraphics[width=75mm]{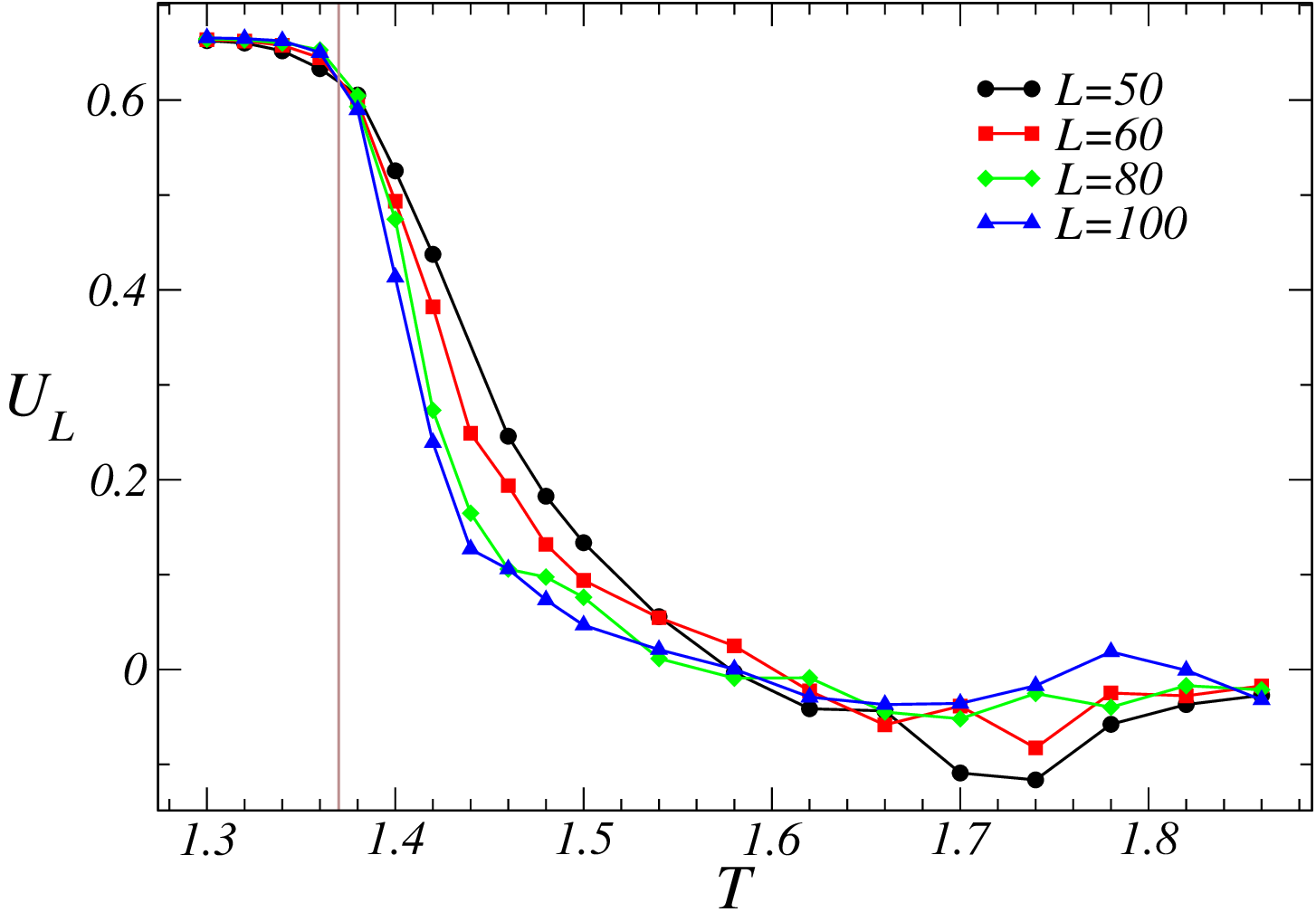}\label{fignewa}}
\subfigure[]{\includegraphics[width=77mm]{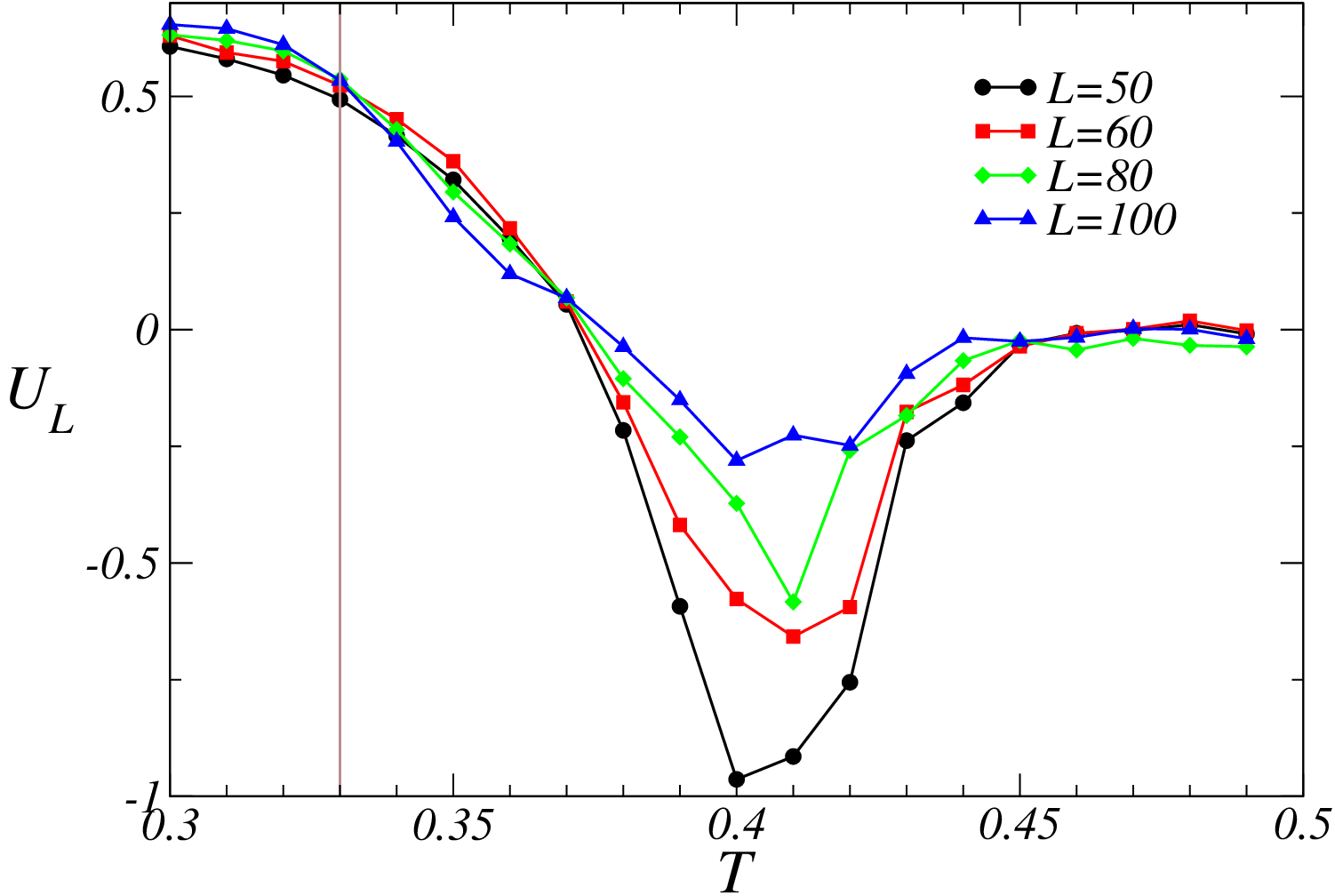}\label{fignewb}}
\vspace{-3mm}
\caption{Binder cumulant $U_L$ as a function of the temperature $T$ for different system sizes. The frequency is $\omega= 0.04\pi$ and $J=1$.
(a) $h_0=1.0$, where the crossing point gives $T_c=1.37(1)$. (b)  $h_0=2.8$, where the crossing point gives $T_c=0.33(1)$. The minimum 
in (b) that decreases with system size is not due to a first-order phase transition but rather a finite-size effect \cite{korn02}.
}
\label{fignew}
\end{figure}

For the two-dimensional Ising model with nearest neighbors interactions, a configuration $\mathbf{s}=(s_1,s_2,\ldots,s_{L^2})$ at time $t$ has energy 
\begin{equation}
E_{\mathbf{s}}(t)\equiv -J\sum_{ \langle ij\rangle} s_is_j-h(t)\sum_i{s_i},
\end{equation}
where the first sum is over nearest neighbors, $h(t)$ is given by (\ref{eqfield}), and we consider periodic boundary conditions.
The Binder cumulant is defined as \cite{land14}
\begin{equation}
U_L\equiv1-\langle M^4\rangle/(3\langle M^2\rangle^2),
\end{equation}
where $M$ represents the magnetization integrated over a period and the brackets denote an average over stochastic trajectories. 
We calculated this Binder Cumulant with numerical simulations, which are explained below. The critical temperatures are determined 
from the crossing points of the Binder Cumulant in Fig. \ref{fignew}. The lack of a minimum of the Binder cumulant that crosses from 2/3 to 0 in Fig. \ref{fignewa} is 
an indicator of a second-order phase transition. The minimum of the Binder cumulant at a negative value in Fig. \ref{fignewb} is an indicator of a first-order 
phase transition, and this transition has been interpreted to be first-order from this kind of numerical 
result \cite{chak99}. However, this minimum has been shown to be finite size effect with a solid theoretical argument and extensive 
numerical simulations that show that the minimum disappears for large enough systems  \cite{korn02} (see Fig. \ref{fignewb}). The basic idea of the 
theoretical argument is that for small systems the temperature at which the system
crossover from the multiple droplet to the single droplet regime is above the critical temperature, leading to major differences in the 
probability distribution of the order parameter \cite{korn02}. Hence, convincing numerical evidence supports that in two dimensions the transition
is always second-order, independent of $h_0$ and $\omega$. Our numerical results in Fig. \ref{fignewb} indicate a minimum that decreases with system size, in agreement 
with the results from \cite{korn02}.

The system reaches a periodic steady state characterized by the probability $P_\mathbf{s}(t)$, which has a period $2\pi/\omega$.
The entropy production rate per spin in this periodic steady state is defined as \cite{seif12} 
\begin{equation}
\sigma_L\equiv \frac{1}{L^2}\frac{\omega}{2\pi}\int_0^{2\pi/\omega}dt\left(\sum_{\mathbf{s}\mathbf{s}'}w_{\mathbf{s}\mathbf{s}'}(t)P_\mathbf{s}(t)\frac{E_{\mathbf{s}}(t)-E_{\mathbf{s}'}(t) }{T}\right),     
\end{equation}
where $w_{\mathbf{s}\mathbf{s}'}(t)$ is the transition rate from  state $\mathbf{s}$ to state $\mathbf{s}'$ at time $t$. These transition rates are nonzero only if the configurations $\mathbf{s}$ and 
$\mathbf{s}'$ differ by one spin flip and they fulfill the detailed balance relation
$w_{\mathbf{s}\mathbf{s}'}(t)/w_{\mathbf{s}'\mathbf{s}}(t)=\textrm{e}^{[E_{\mathbf{s}}(t)-E_{\mathbf{s}'}(t)]/T}$. The factor $L^{-2}$ makes $\sigma_L$ finite in the thermodynamic limit, where
\begin{equation}
\sigma\equiv\lim_{L\to \infty}\sigma_L.
\end{equation}
The rate of dissipated heat is $\dot{q}=T\sigma$, which is equal to the rate of work done on the system due to the first law.

Numerical simulations were performed with the following procedure. The initial condition is 
a random configuration of spins, corresponding to $T\to \infty$. The time $t$ is discretized with the integer variable $n$, in such a way 
that for $n=L^2$ we have $t=1$, i.e., the time $t$ is in units of Monte Carlo steps. We use the standard 
metropolis rule for flipping a spin \cite{newm99}. A randomly chosen spin $s_i$ may flip depending on the energy difference
\begin{equation}
\Delta E_i(n/L^2)= 2s_i[J\sum_js_j+h(n/L^2)],
\end{equation}    
where the sum in $j$ is over the four nearest neighbors. If this energy difference is negative the spin flips with probability one, and if it
 is positive, the spin flips with probability $\textrm{e}^{-\Delta E_i(n/L^2)/T}$.

The entropy production rate $\sigma_L$ was computed in the following way. After a 
certain transient the system reaches a periodic steady state and we compute the change in the entropy of the external medium 
$\Delta S$ from time $0$, after the transient, to time $\mathcal{T}$, which corresponds 
to several periods $2\pi/\omega$. If the system jumps from a configuration $\mathbf{s}$ to a configuration $\mathbf{s}'$ the entropy $\Delta S$ changes by an amount 
$[E_{\mathbf{s}}(t)-E_{\mathbf{s}'}(t)]/T$. The entropy production rate per spin is then given by
\begin{equation}
\sigma_L=\Delta S/(\mathcal{T}L^2).
\label{entsim}
\end{equation}

\subsection{Critical behavior of the entropy production}

\begin{figure}
\subfigure[]{\includegraphics[width=75mm]{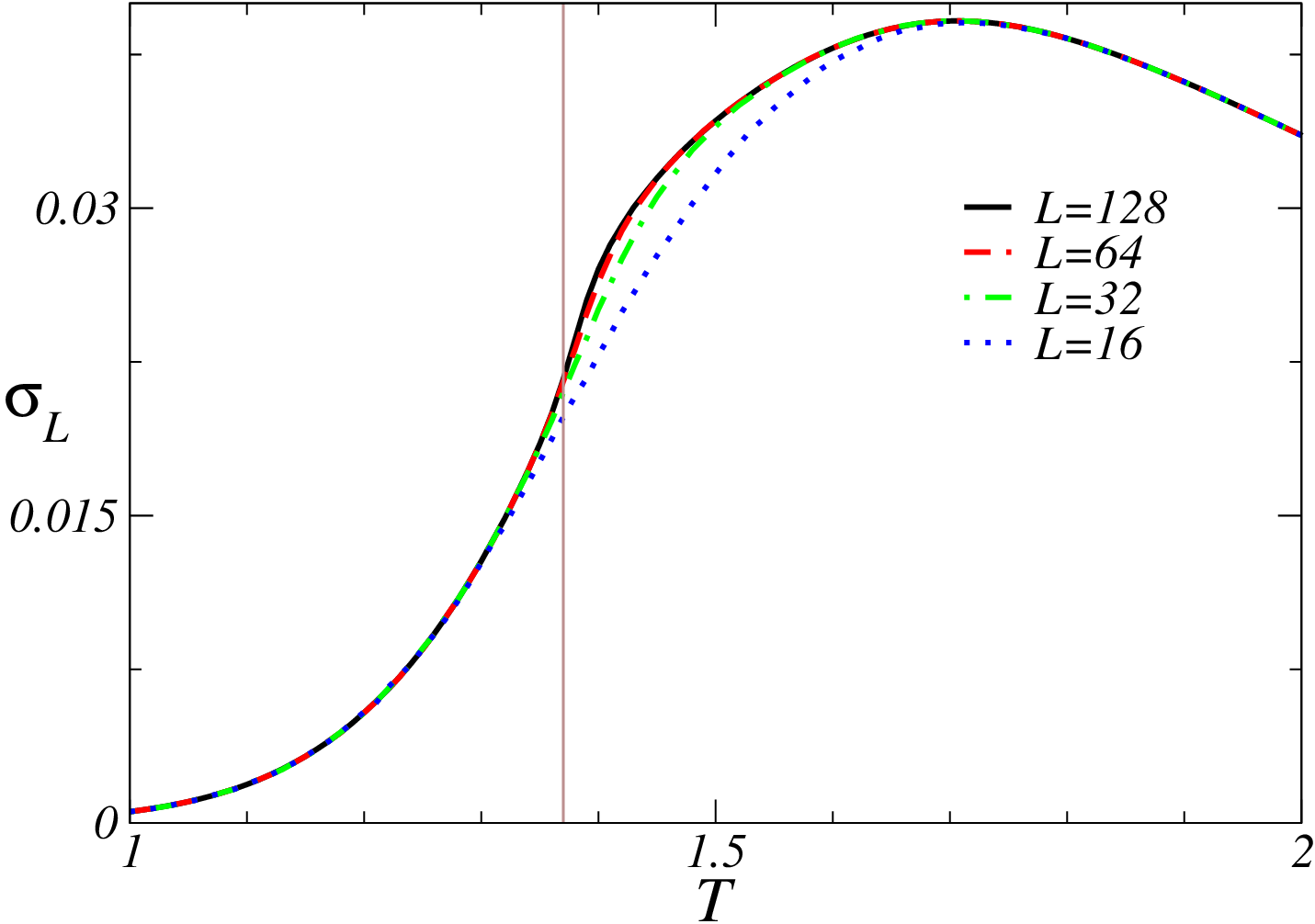}\label{fig3a}}
\subfigure[]{\includegraphics[width=77mm]{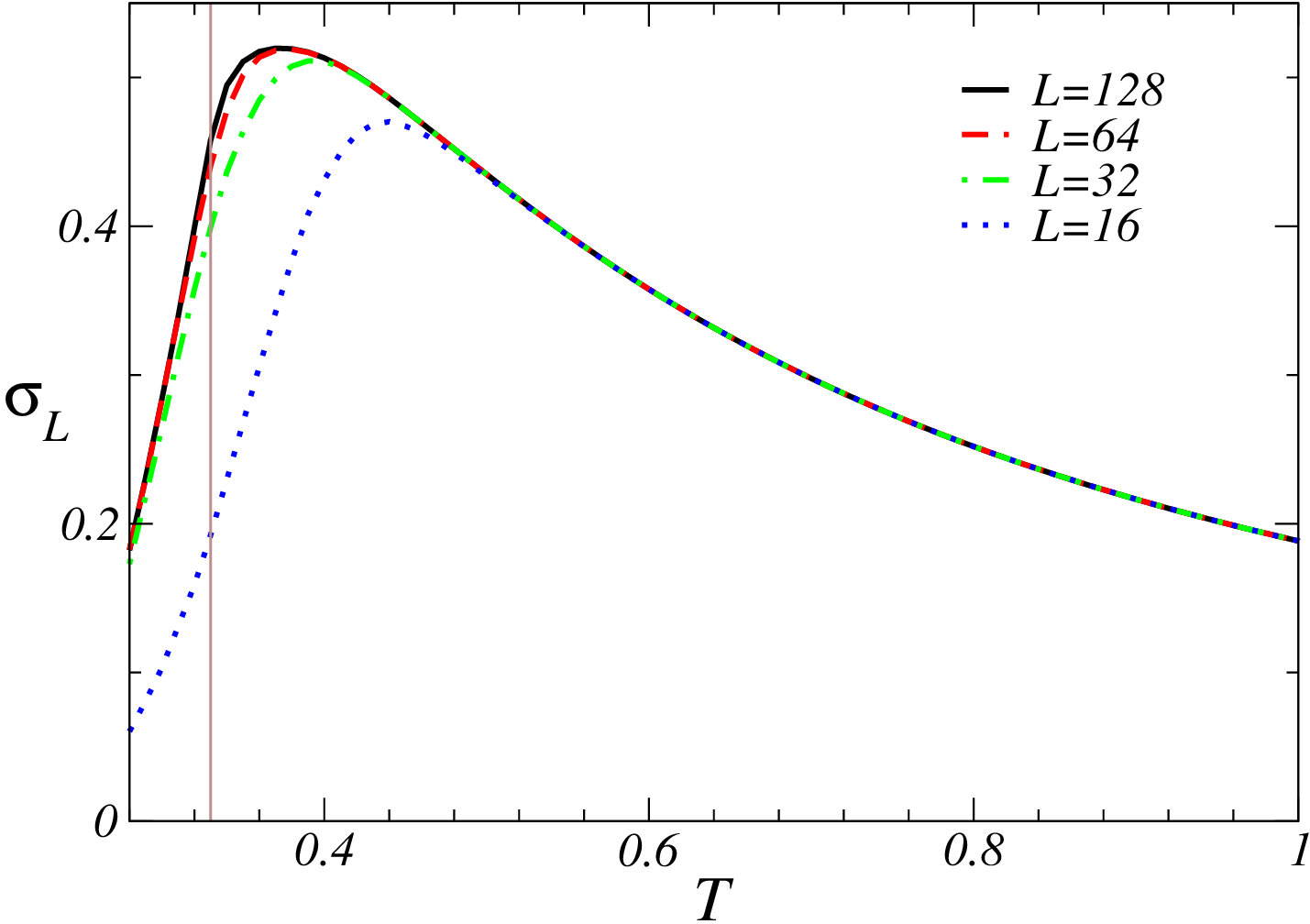}\label{fig3b}}
\vspace{-3mm}
\caption{Entropy production rate $\sigma_L$ as a function of the temperature $T$.  The frequency is $\omega= 0.04\pi$ and $J=1$.
(a) $h_0=1.0$ and (b)  $h_0=2.8$. The vertical lines indicate the critical temperatures given in the caption of Fig. \ref{fignew}.
}
\label{fig3}
\end{figure}

In Fig. \ref{fig3} we plot the entropy production rate $\sigma_L$ as a function of the temperature $T$ for two different values of $h_0$. In both cases the entropy production rate
has a maximum above $T_c$. At the critical point, the entropy production rate seems to have an inflection, indicating a divergence 
of the first derivative of $\sigma_L$ at the critical point, i.e.,
\begin{equation}
d_T\sigma\sim |T-T_c|^{-\alpha}.
\end{equation}

\begin{figure}
\subfigure[]{\includegraphics[width=75mm]{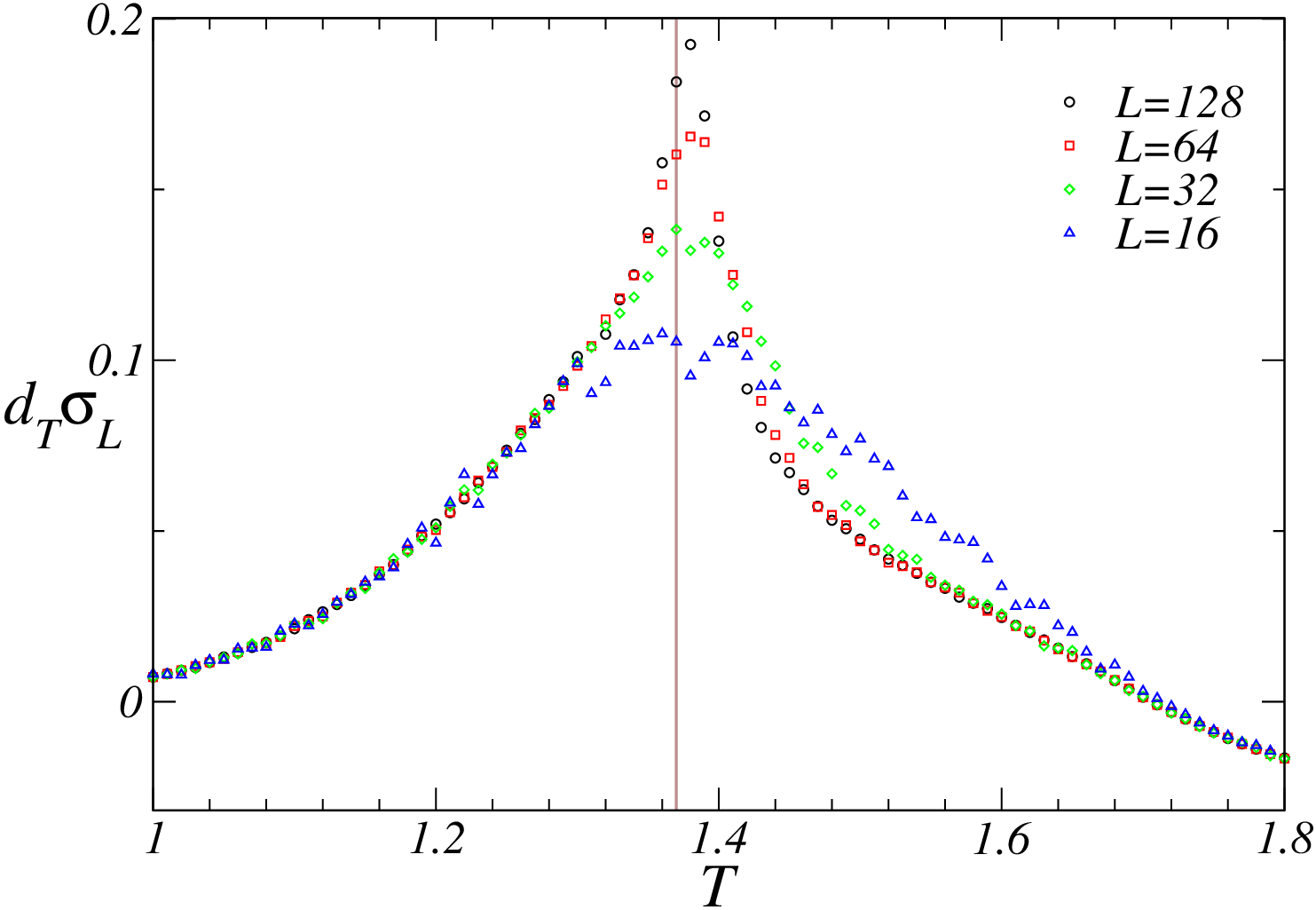}\label{fig4a}}
\subfigure[]{\includegraphics[width=75mm]{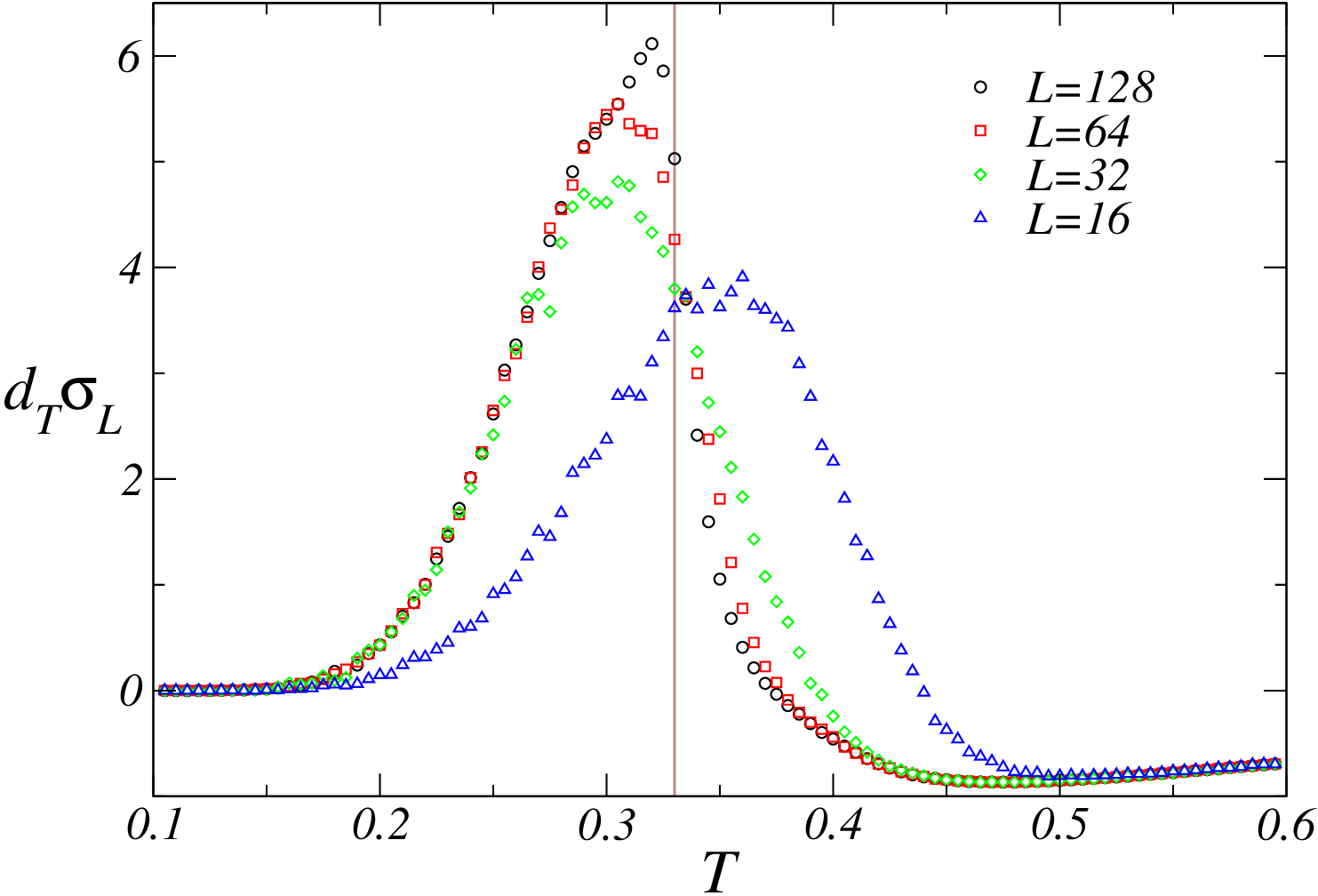}\label{fig4b}}
\vspace{-3mm}
\caption{The first derivative $d_T\sigma_L$ as a function of the temperature $T$.  The frequency is $\omega= 0.04\pi$ and $J=1$.
(a) $h_0=1.0$ and (b) $h_0=2.8$. The vertical lines indicate the critical points given in the caption of Fig. \ref{fignew}.
}
\label{fig4}
\end{figure}

\begin{figure}
\subfigure[]{\includegraphics[width=75mm]{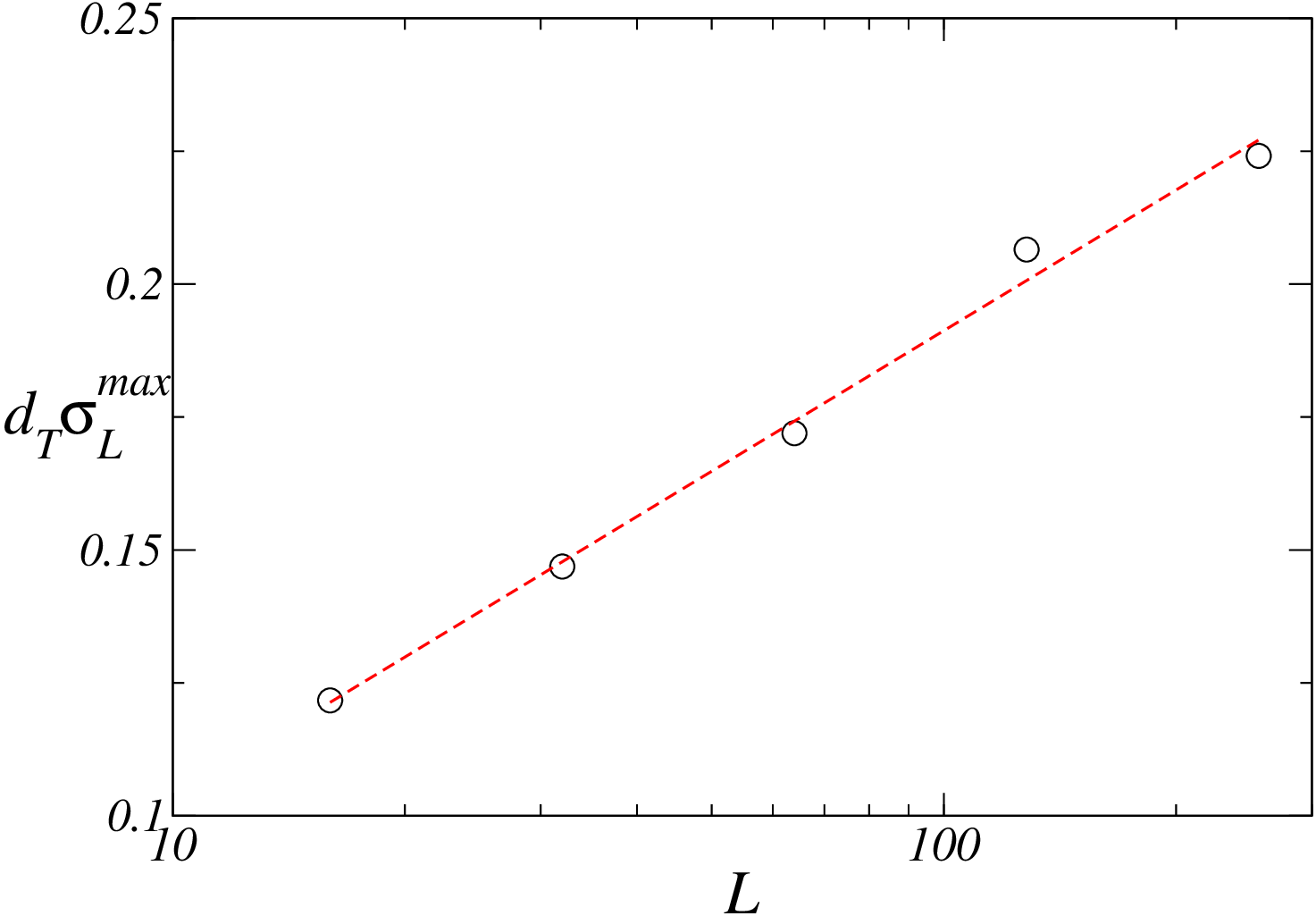}\label{fig5a}}
\subfigure[]{\includegraphics[width=75mm]{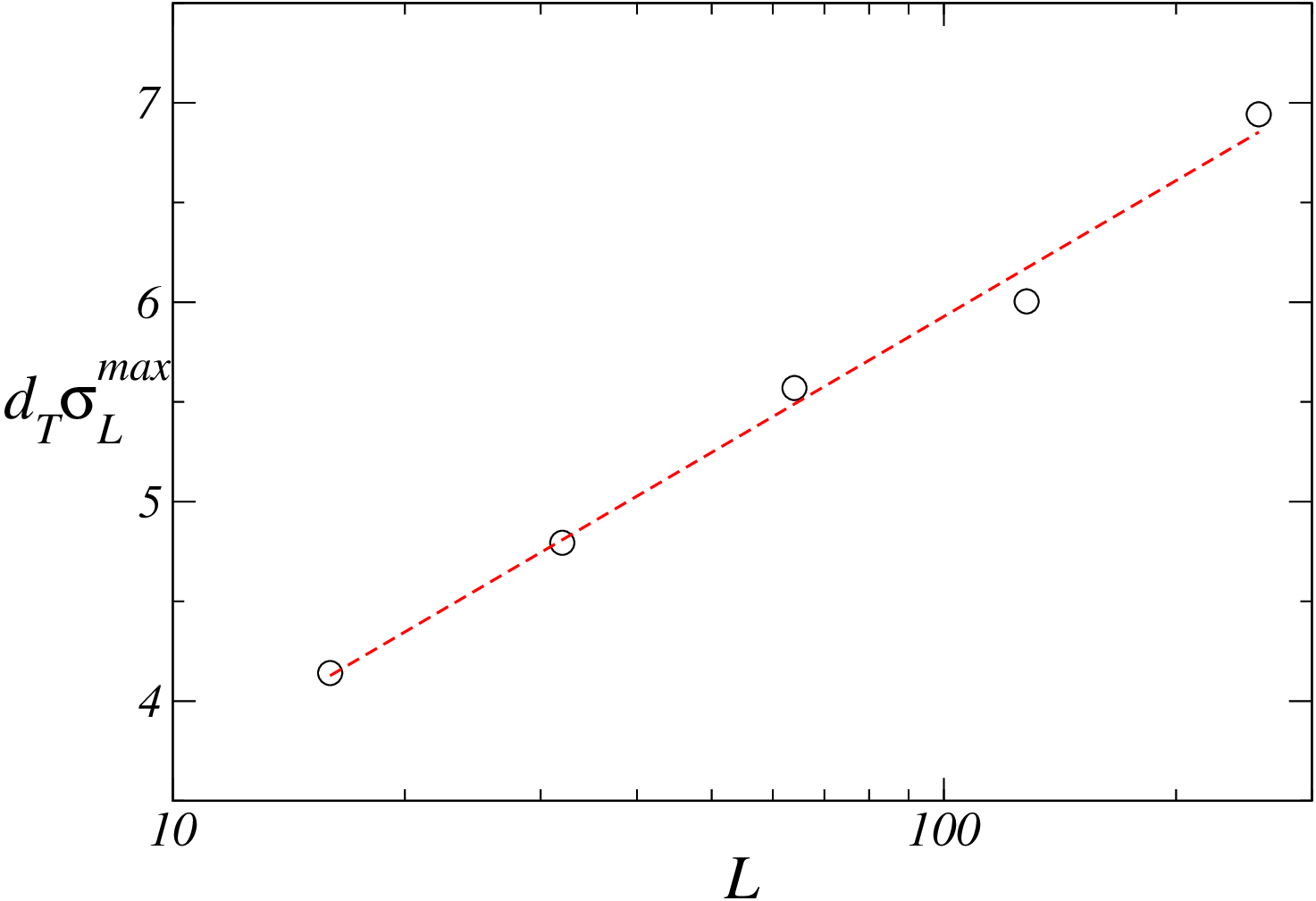}\label{fig5b}}
\vspace{-3mm}
\caption{Maximum of the first derivative $d_T\sigma_L^{\textrm{max}}$ as a function of the system size $L$.  The frequency is $\omega= 0.04\pi$ and $J=1$.
(a) $h_0=1.0$, where the transition is second-order. (b) $h_0=2.8$, where the transition is first-order.
In both cases this maximum behaves as $d_T\sigma_L^{max}\sim \ln L$. 
}
\label{fig5}
\end{figure}

Direct numerical evaluation of the exponent $\alpha$ with off-critical simulations turn out to be difficult. As shown in Fig. \ref{fig4},  the 
first derivative $d_T\sigma_L$ for a finite system has a maximum that increases with $L$, in agreement with the expectation that $d_T\sigma$ diverges at criticality. 
Plotting this maximum  $d_T\sigma_L^{\textrm{max}}$ as a function of 
system size in Fig \ref{fig5}, we obtain  
\begin{equation}
d_T\sigma_L^{\textrm{max}}\sim \ln L.
\label{entcri}
\end{equation}
The exponent $\alpha$ is related to an exponent $\psi$, defined by $d_T\sigma_L^{\textrm{max}}\sim L^{\psi}$, through the scaling relation $\alpha=\psi\nu$, where 
$\nu$ is the critical exponent characterizing the divergence of the correlation length at criticality ($\nu\simeq 1.1$ \cite{chak99}). Relation (\ref{entcri}) implies $\psi=0$, and, therefore,
$\alpha=0$. 

The critical behavior of the entropy production rate is the same in both plots shown in Fig. \ref{fig5}. This result provides further support for a 
second-order phase transition also for the amplitude $h_0=2.8$, for which the Binder cumulant for small systems shows 
a minimum in Fig \ref{fignewb}, in the sense that with a first-order phase transition the system would explore different regions of the 
phase-space, which could lead to a discontinuity in the entropy production at $T_c$. However, the lack of a jump in $\sigma$ cannot be taken 
as a demonstration that the transition is not first-order: a relation connecting the average magnetization integrated over a period with the 
entropy production is not known, and, therefore, a discontinuity in the order parameter does not necessarily imply a discontinuity in the entropy production.

The same kind of critical behavior of the first derivative of the entropy production, characterized by a logarithmic divergence, 
has been observed in a majority vote model \cite{croc05}  and in a Ising model in contact with two heat baths \cite{tome12}.

\section{Critical behavior of the learning rate}
\label{sec4}
\subsection{Stochastic external field}

We now consider a two-dimensional Ising model with a magnetic field that changes at stochastic times. A similar model has been considered in \cite{acha98}.  
This magnetic field changes at a rate $\Gamma$ between the values $h_0$ and $-h_0$. The system and 
external field together form a bipartite Markov process \cite{hart14,horo14}, which has $2\times 2^{L^2}$ states.

A state of this bipartite process is characterized by the vector $\mathbf{s}$ and a binary variable $x=\pm 1$ that 
indicates whether the external field is $h_0$ or $-h_0$. The transition rate from an state $(\mathbf{s},x)$ to a
state $(\mathbf{s}',x')$ is
\begin{equation}
w_{\mathbf{s}\mathbf{s}'}^{xx'}\equiv 
\cases{
w^{xx'}_{\mathbf{s}}=\Gamma\qquad\textrm{if $x\neq x'$ and $\mathbf{s}= \mathbf{s}'$ }\\
w^{x}_{\mathbf{s}\mathbf{s}'}=\chi_{\mathbf{s}\mathbf{s}'}/(1+\textrm{e}^{(E^x_{\mathbf{s}'}-E^x_\mathbf{s})/T})\qquad\textrm{if $x= x'$ and $\mathbf{s}\neq \mathbf{s}'$ }\\
0 \qquad\textrm{if $x\neq x'$ and $\mathbf{s}\neq \mathbf{s}'$,}
}
\end{equation}
where $\chi_{\mathbf{s}\mathbf{s}'}$ is $1$ if both configurations differ by a single spin flip and $0$ otherwise. The energy $E^x_\mathbf{s}$ is given by
\begin{equation}
E^x_\mathbf{s}\equiv  -J\sum_{ \langle ij\rangle} s_is_j-xh_0\sum_i{s_i}.
\end{equation}

A bipartite Markov process has two kinds of jumps, internal ones that lead to a spin flip and external ones that change the external field. 
The dissipated heat is related to the internal jumps, whereas the external jumps are related to work. Hence, only internal jumps appear 
in the entropy production rate per spin $\sigma_L$, which is defined as \cite{seif12}
\begin{equation}
\sigma_L\equiv\frac{1}{L^2}\sum_x\sum_{\mathbf{s}\mathbf{s}'}P_\mathbf{s}^{x}w^x_{\mathbf{s}\mathbf{s}'}\frac{(E^x_\mathbf{s}-E^x_{\mathbf{s}'})}{T},
\end{equation}
where $P_\mathbf{s}^{x}$ is the stationary distribution.

We have performed continuous-time Monte Carlo simulations of this model, using a method related to the method introduced in \cite{bort75}. The 
main difference is that we also have to account for jumps that lead to a change in the magnetic field. In our algorithm, at each jump, 
there is a probability $1-p^{spin}$, which depends on the state of the system, that the magnetic field changes. A spin flip happens with
probability $p^{spin}$, and is executed with the procedure explained in \cite{bort75}. 

The parameter $\Gamma$ is written as $\Gamma=\gamma L^2$, where for different system sizes  $\gamma$ is kept fixed. The 
parameter $\Gamma$ must scale as $L^2$ for the following reason. The escape rate 
of an state $(\mathbf{s},x)$ is $r_{\mathbf{s}x}=r^{spin}_{\mathbf{s}x}+\Gamma$, where $r^{spin}_{\mathbf{s}x}$ is the sum of all 
transition rates that lead to a spin flip. Since there are $L^2$ spins that can be flipped, the parameter $\Gamma$ has to scale with $L^2$ for the probability of
a change in the field $1-p^{spin}=\Gamma/(r^{spin}_{\mathbf{s}x}+\Gamma)$ to be conserved with a change in system size.  

The entropy production is calculated  by adding 
$(E^x_\mathbf{s}-E^x_{\mathbf{s}'})/T$ to $\Delta S$ every time a jump from $(x,\mathbf{s})$ to $(x,\mathbf{s}')$ occurs. If the simulation runs 
for a time $\mathcal{T}$, after some transient, the entropy production rate per  spin is calculated with expression (\ref{entsim}).

\begin{figure}
\subfigure[]{\includegraphics[width=75mm]{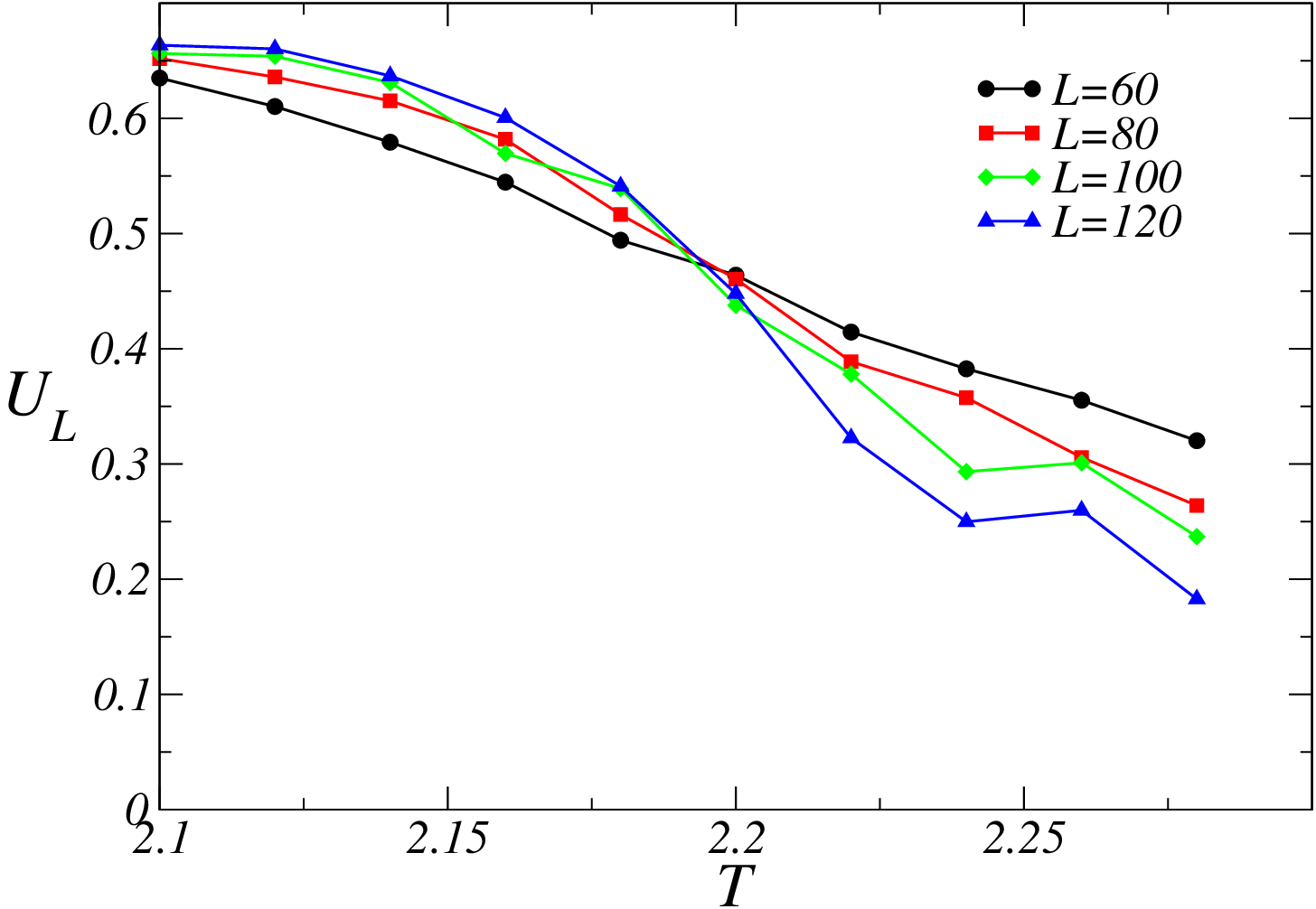}\label{fig6a}}
\subfigure[]{\includegraphics[width=75mm]{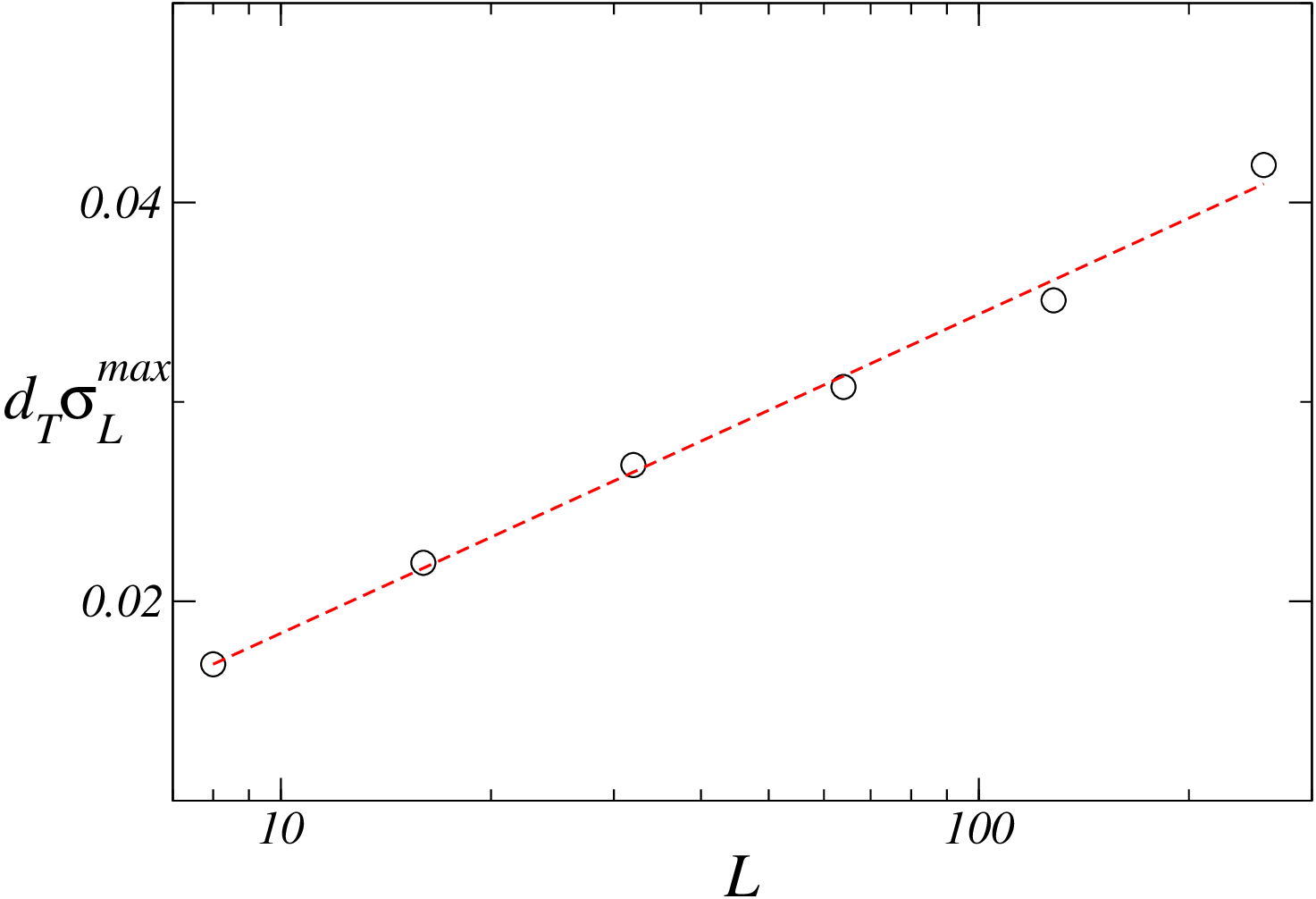}\label{fig6b}}
\vspace{-3mm}
\caption{Critical behavior of $\sigma$ for the model with a stochastic field.  The parameters are $\gamma=0.002$, $J=1$, and $h_0=0.5$.
(a) Binder Cumulant for different systems sizes indicating $T_c=2.19(1)$. (b) Maximum of the first derivative of the entropy production rate $d_T\sigma_L^{\textrm{max}}$
as a function of $L$. 
}
\label{fig6}
\end{figure}

The critical point is again determined with the Binder cumulant 
\begin{equation}
U_L=1-\langle m^4\rangle/(3\langle m^2\rangle^2),
\end{equation}
where the brackets denote an average over stochastic trajectories and $m= \sum_is_i/L^2$.  
The Binder cumulant in Fig. \ref{fig6a} indicates a second-order phase transition. The critical behavior 
of the entropy production rate is the same as in the previous model, as shown in Fig. \ref{fig6b}, where the maximum 
of the first derivative of the entropy production rate follows the behavior in Eq. (\ref{entcri}).  

\subsection{Learning rate}

The Ising model with a stochastic field allows us to consider a further aspect related to information theory.
The external field can be interpreted as a stochastic signal and the system of spins as a sensor that  follows the signal.
It turns out that there is a quantity, called learning rate \cite{bara14a,hart16}, that characterizes the rate at which the system obtains information 
about the signal. Technically, this learning rate is a time derivative of the mutual information between system and external field.

In the stationary state, the learning rate (per spin) $\lambda_L$ is given by \cite{bara14a,hart16}
\begin{equation}
\lambda_L= \frac{\Gamma}{L^2}\sum_{\mathbf{s}}(P^{+1}_\mathbf{s}-P^{-1}_\mathbf{s})\ln\frac{P^{+1}_\mathbf{s}}{P^{-1}_\mathbf{s}}.
\end{equation}
The second law for a sensor following a signal reads \cite{hart14,horo14,bara14a}
\begin{equation}
\lambda_L\le \sigma_L.
\label{inelearn}
\end{equation} 
The learning rate $\lambda_L$ that quantifies how much information the system obtains about the signal is bounded by $\sigma_L$, which quantifies heat dissipation.
This inequality allows for the definition of the informational efficiency $\eta= \lambda_L/\sigma_L$. 

\begin{figure}
\subfigure[]{\includegraphics[width=75mm]{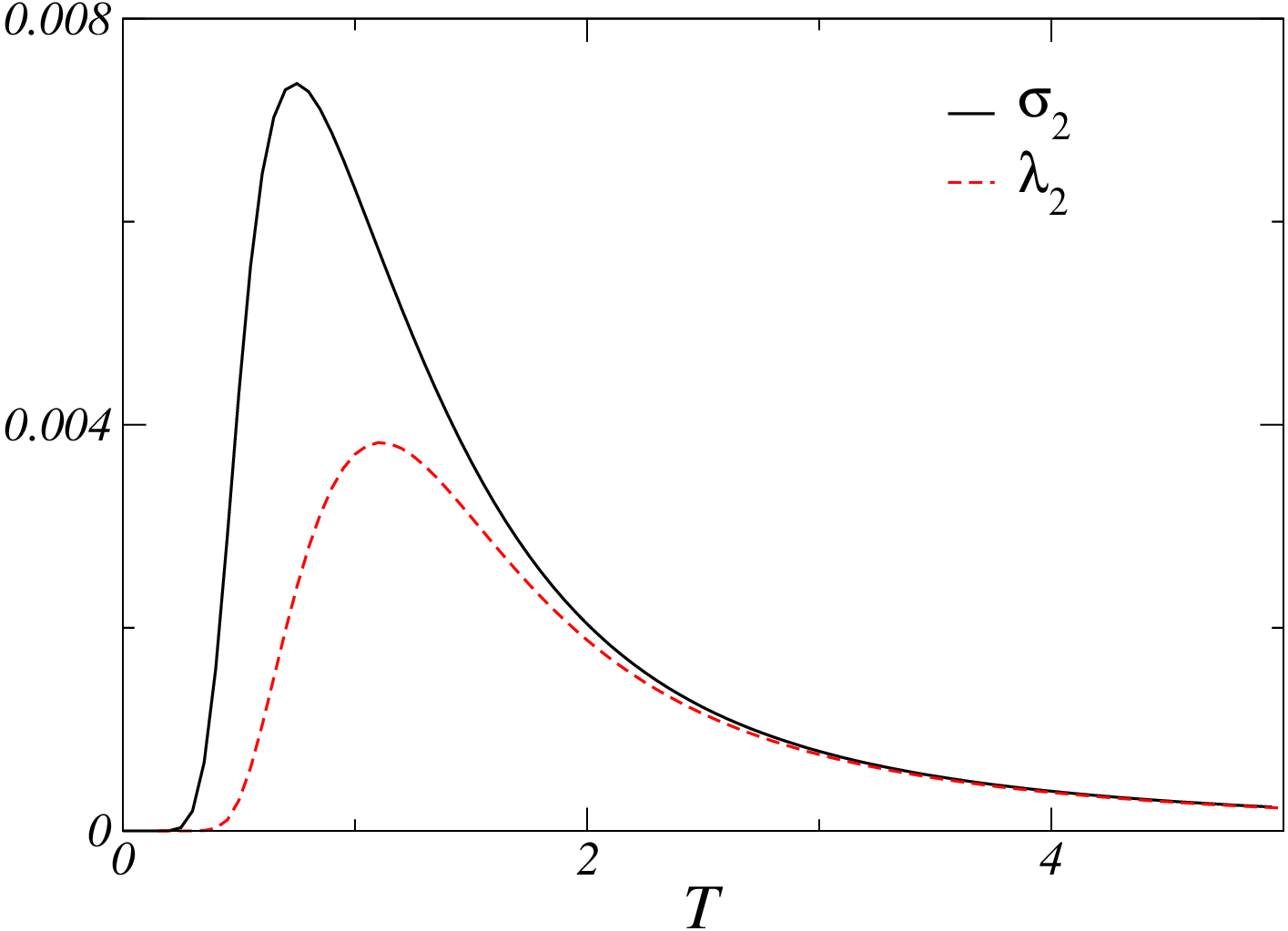}\label{fig7a}}
\subfigure[]{\includegraphics[width=75mm]{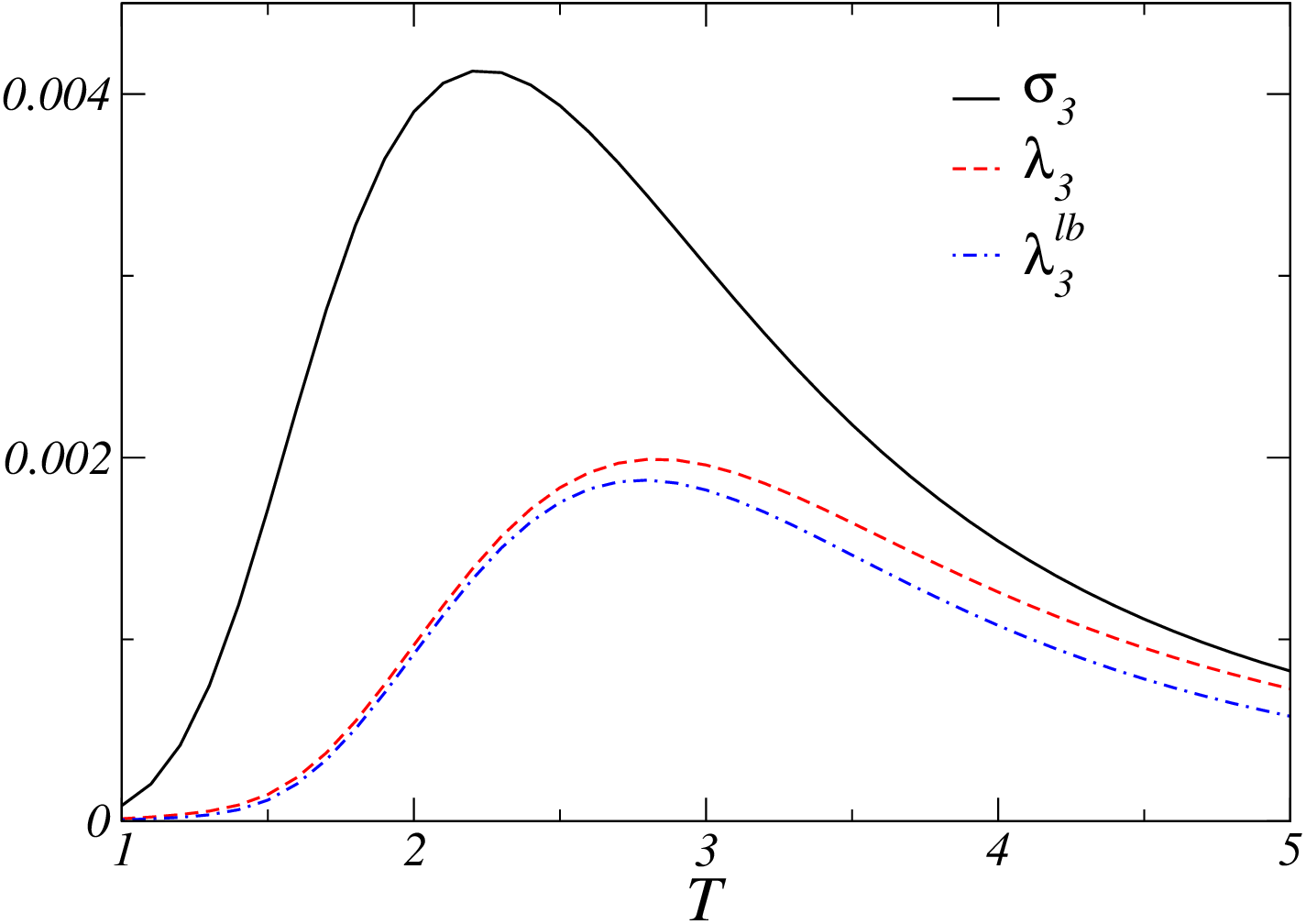}\label{fig7b}}
\vspace{-3mm}
\caption{Exact learning rate $\lambda_L$ and entropy production rate $\sigma_L$ for small systems.  The parameters are $\gamma=0.002$, $J=1$, and $h_0=0.5$.
(a) $L=2$. (b) $L=3$. 
}
\label{fig7}
\end{figure}

In Fig. \ref{fig7} we plot the learning rate $\lambda_L$ and the entropy production rate $\sigma_L$ as functions of temperature $T$ for $L=2,3$. They both 
have maxima at some intermediate values of $T$. These figures were obtained with the exact calculation of the eigenvector of 
the stochastic matrix that is associated with the eigenvalue $0$, which lead  to the stationary distribution $P^x_{\mathbf{s}}$.

For larger values of $L$ we have to use Monte Carlo simulations. The problem that arises for the calculation of $\lambda_L$ in  simulations is that
the increment in the learning rate after a jump depends on the nonequilibrium stationary probability $P^{x}_{\mathbf{s}}$, which is not known. 

We propose the following lower bound on the learning rate. Instead of the microscopic configuration $\mathbf{s}$ we consider some mesoscopic variable $a$. In particular,
we consider a variable that gives the ``class'' of the spin in the continuous-time simulation \cite{bort75}. This variable takes the orientation of a spin and its nearest neighbors 
into account and has $10$ possible outcomes: for each spin orientation the number of nearest neighbors with $s_j=1$ can go from $0$ to $4$. The lower bound is then written as
\begin{equation}
\lambda_L^{lb}\equiv  \frac{\Gamma}{L^2}\sum_{a}(P^{+1}_a-P^{-1}_a)\ln\frac{P^{+1}_a}{P^{-1}_a}.
\label{learnlb}
\end{equation}
This lower bound fulfills $\lambda_L^{lb}\le \lambda_L$, which is illustrated in Fig. \ref{fig7b},  due to the log sum inequality \cite{cove06}. 
The probability $P^{x}_a$ can be calculated in a Monte Carlo simulation by calculating the density of spins in each one of the ten classes in the steady state.

\begin{figure}
\centering\includegraphics[width=99mm]{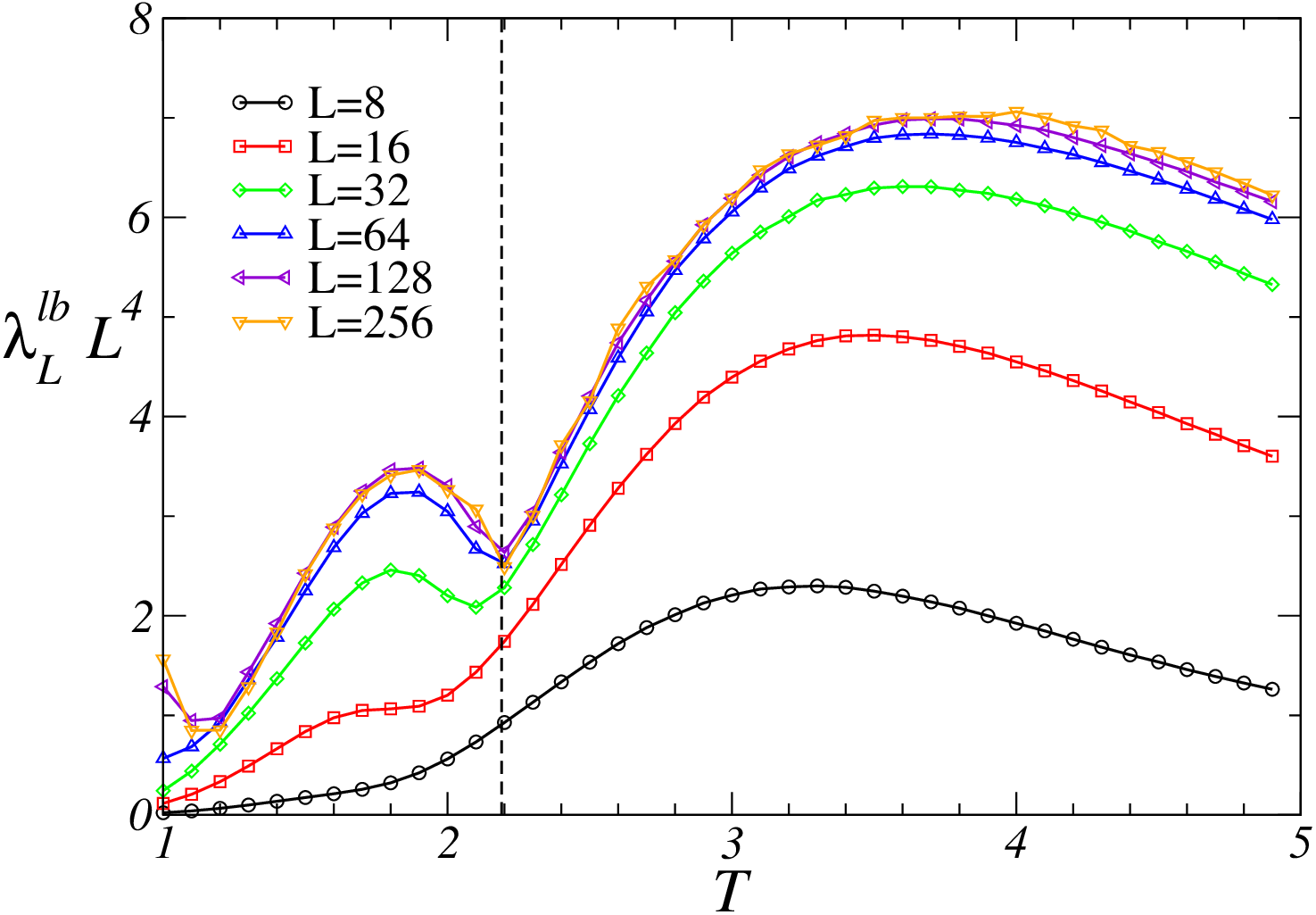}
\vspace{-3mm}
\caption{Critical behavior of the lower bound on the learning rate $\lambda_L^{lb}$.  The parameters are $\gamma=0.002$, $J=1$, and $h_0=0.5$. The critical temperature $T_c=2.19(1)$ 
is indicated by the dotted vertical line.}
\label{fig8}
\end{figure}

It turns out that the lower bound $\lambda_L^{lb}$ scales as $L^{-4}$, going to zero in the limit of infinite system size. 
This scaling can be understood with the following heuristic argument. The probability of changing the magnetic field in a transition, instead of 
flipping a spin, is conserved as $L$ is increased. Therefore, after a change in the field, the average number of spin flips 
before the next change in the external field is conserved for increasing $L$. However, the number of spins flips to equilibrate the system 
is proportional to $L^2$. As the number of spin flips is constant and the number of spin flips necessary for equilibration scale as $L^2$, it
is reasonable to expect that the difference $|P^{+1}_a-P^{-1}_a|\sim L^{-2}$, which we have confirmed numerically. 
From the expression (\ref{learnlb}) we obtain $\lambda_L^{lb}\sim L^{-4}$.

In Fig. \ref{fig8} we plot the scaled learning rate $\lambda_L^{lb}L^{4}$ as a function of the temperature $T$. 
For large $L$, this lower bound seems to have a kink, in the form of a local minimum, at the critical point. Hence, our results indicated that the
first derivative of $\lambda_L^{lb}$ with respect to $T$ is discontinuous in the limit $L\to \infty$. 
The critical behavior of the lower bound on the learning rate is then different from that of the entropy production. 
We note that the lower bound on the efficiency  $\lambda_L^{lb}/\sigma_L\le\eta$ goes to zero in the thermodynamic limit due to the different scaling of  
$\lambda_L^{lb}$ in relation to $\sigma_L$.

\section{Conclusion}
\label{sec5}

We have analyzed the critical behavior of the entropy production rate of a nonequilibrium Ising 
model subjected to a time-dependent periodic field. For the mean-field model, this entropy production rate
is found to have a jump at criticality if the transition is first-order. However, if the transition is second-order, the entropy production rate
is continuous but its first derivative is discontinuous at criticality. For the two-dimensional model, for which the transition is second-order, the first 
derivative of the entropy production rate has a logarithmic divergence at the critical point. The novelty of our results in relation to previous studies
on the critical behavior of entropy production rate \cite{croc05,tome12,deol11,bara12} are the following. The models analyzed here are driven
by an external periodic protocol, in contrast to previous studies that consider models driven by 
a fixed thermodynamic force; the entropy production rate was found to have a jump at criticality for 
the mean-field model in the region with a first-order phase transition, which is a critical behavior 
that has not been observed in previous studies. Furthermore, the results for the deterministic field support
the lack of a first-order phase transition in two-dimensions. 

We have also investigated the critical behavior of the learning rate for the model with 
an external field that changes at stochastic time-intervals between two values. It turns 
out that the calculation of the learning rate within numerical simulations would require 
the unknown nonequilibrium stationary distribution. We introduced a lower bound 
on the learning rate that can be calculated within numerical simulations. Our numerics 
indicates that the critical behavior of this lower bound is different from the one of the
entropy production: it has a local minimum at the critical point and its first 
derivative seems to be discontinuous.

Our results on the Ising model with a stochastic field offers two fresh perspectives. First,
most studies on the relation between information and thermodynamics consider small systems.
However, the inequalities for bipartite processes \cite{hart14,horo14} are also valid for macroscopic
systems with a large number of states, as explicitly illustrated here. For example, it would 
be interesting to build a model of a macroscopic Maxwell's demon using the framework for bipartite systems.
Second, this model allows for the definition of an informational efficiency. However, 
the lower bound on the efficiency, which is the quantity we could calculate, turned out to go to zero in 
the thermodynamic limit. Analyzing the critical behavior of efficiency in nonequilibrium models is also 
an interesting perspective. 

This work and the few previous studies on the critical behavior of entropy production  demonstrate that the 
average entropy production can be a useful observable to determine the 
critical point of a generic nonequilibrium phase transition. As an interesting direction for future work, higher 
order moments of the fluctuating entropy production could be even more effective for a precise determination of the critical point.
While the entropy production has been found to display three distinct behaviors at the critical line, i.e., a logarithmic divergence on its first derivative, 
a discontinuity on its first derivative, and a discontinuity, the deeper question whether entropy production can be used to classify nonequilibrium phase 
transitions in a meaningful way remains open.


{\noindent \textbf{Acknowledgements}}\newline We thank Shamik Gupta
for carefully reading the manuscript and Per Arne Rikvold for pointing out \cite{korn02}.

\section*{References}


\end{document}